\documentclass[12pt,preprint]{aastex}  

\usepackage{graphicx}
\usepackage{txfonts}
\newcommand{\va}{v_{\mathrm{A}}}
\newcommand{\vap}{v_{\mathrm{A}f}}
\newcommand{\vac}{v_{\mathrm{A}c}}

\newcommand{\pd}{\partial}
\newcommand{\td}{\tau_{\mathrm{D}}}
\newcommand{\tdp}{\tau_{\mathrm{D}} / P}

\newcommand{\etac}{\eta_{\rm C}}
\newcommand{\etah}{\eta_{\rm H}}
\newcommand{\etat}{\tilde{\eta}}
\newcommand{\etact}{{\tilde{\eta}_{\rm C}}}
\newcommand{\etaht}{\tilde{\eta}_{\rm H}}

\newcommand{\etactf}{\tilde{\eta}_{{\rm C}f}}
\newcommand{\etactc}{\tilde{\eta}_{{\rm C}c}}
\newcommand{\mut}{\tilde{\mu}}

\begin{document}

	\title{RESONANTLY DAMPED KINK MAGNETOHYDRODYNAMIC WAVES IN A PARTIALLY IONIZED FILAMENT THREAD}

	\shorttitle{RESONANTLY DAMPED KINK MHD WAVES IN A PARTIALLY IONIZED FILAMENT THREAD}

   \author{R. Soler, R. Oliver, and J. L. Ballester}
   \affil{Departament de F\'isica, Universitat de les Illes Balears,
              E-07122, Palma de Mallorca, Spain}
              \email{[roberto.soler;ramon.oliver;joseluis.ballester]@uib.es}

  \begin{abstract}
Transverse oscillations of solar filament and prominence threads have been frequently reported. These oscillations have the common features of being of short period (2--10~min) and being damped after a few periods. Kink magnetohydrodynamic (MHD) wave modes have been proposed as responsible for the observed oscillations, whereas resonant absorption in the Alfv\'en continuum and ion-neutral collisions are the best candidates to be the damping mechanisms. Here, we study both analytically and numerically the time damping of kink MHD waves in a cylindrical, partially ionized filament thread embedded in a coronal environment. The thread model is composed of a straight and thin, homogeneous filament plasma, with a transverse inhomogeneous transitional layer where the plasma physical properties vary continuously from filament to coronal conditions. The magnetic field is homogeneous and parallel to the thread axis. We find that the kink mode is efficiently damped by resonant absorption for typical wavelengths of filament oscillations, the damping times being compatible with the observations. Partial ionization does not affect the process of resonant absorption, and the filament plasma ionization degree is only important for the damping for wavelengths much shorter than those observed. To our knowledge, this is the first time that the phenomenon of resonant absorption is studied in a partially ionized plasma. 

  \end{abstract}



\section{INTRODUCTION}

The fine-structures of solar prominences and filaments are clearly seen in high-resolution observations. These fine-structures, here called threads, appear as very long ($5'' - 20''$) and thin ($0''.2 - 0''.6$) dark ribbons in H$\alpha$ images of filaments on the solar disk \citep[e.g.,][]{lin04,lin05,lin07,lin08,lin09}, as well as in observations of prominences in the solar limb from the Solar Optical Telescope (SOT) aboard the Hinode satellite \citep[e.g.,][]{okamoto,berger,chae,ning}. Although statistical studies show that the orientation of threads can significantly vary within the same filament \citep{lin04}, vertical threads are more commonly seen in quiescent prominences \citep[e.g.,][]{berger,chae} whereas horizontal threads are usually observed in active region prominences \citep[e.g.,][]{okamoto}. From the theoretical point of view, filament threads have been modeled as magnetic flux tubes anchored in the solar photosphere \citep[e.g.,][]{ballesterpriest,rempel}. In this interpretation, only part of the flux tube would be filled with the cool ($\sim 10^4$~K) filament material, which would correspond to the observed threads. It has been also suggested by differential emission measure studies that each thread might be surrounded by its own prominence-corona transition region (PCTR) where the plasma physical properties would abruptly vary from filament to coronal conditions \citep{cirigliano}. The filament material, roughly composed by 90\% hydrogen and 10\% helium, is only partially ionized for typical filament temperatures, although the precise value of the ionization degree is not well-known and could probably vary in different filaments or even in different threads within the same filament \citep{patsu}.

Oscillations of prominence and filament threads have been frequently reported since telescopes with a high time and spatial resolution became available. Early works by \citet{yi1} and \citet{yi2}, with a relatively low spatial resolution ($\sim 1''$), detected oscillatory variations in Doppler signals and He I intensity from threads in quiescent filaments. Later, H$\alpha$ and Doppler observations with a much better spatial resolution ($\sim 0.''2$) found evidence of oscillations and propagating waves along quiescent filament threads \citep{lin04,lin07,lin09}, while observations from the Hinode spacecraft showed transverse oscillations of threadlike structures in both active region \citep{okamoto} and quiescent \citep{ning} prominences. Common features of these observations are that the reported periods are usually in a narrow range between 2 and 10 minutes, and that they are of small amplitude, with the velocity amplitudes smaller than $\sim 3$~km~s$^{ -1}$. Some theoretical works have attempted to explain these observed oscillations in terms of linear magnetohydrodynamic (MHD) waves supported by the thread body, modeled as a cylindrical magnetic tube \citep{diaz02,dymovaruderman}. These studies concluded that the so-called kink MHD mode is the best candidate to explain transverse, nonaxisymetric thread oscillations, and is also consistent with the reported short periods. An application of this interpretation was performed by \cite{hinode}, who made use of the model by \citet{dymovaruderman} and the observations by \citet{okamoto} to obtain lower limits of the prominence Alfv\'en speed. Similarly, \citet{lin09} interpreted their observations of swaying threads in H$\alpha$ sequencies as propagating kink waves and gave an estimation of the Alfv\'en speed. The reader is refereed to recent reviews by \citet{oliverballester,ballester,banerjee,engvold} for more extensive comments.

Another interesting characteristic of prominence oscillations is that they seem to be damped after a few periods. Although this behavior was previously suggested by the results of some works \citep[e.g.,][]{landman,tsubaki}, it was first extensively investigated by \citet{molowny} and \citet{terradasobs}. These authors studied two-dimensional Doppler time-series from a quiescent prominence and found that oscillations detected in large areas of the prominence were typically damped after 2--3 periods. Similar results were obtained in a more recent work by \citet{mashnich}. This damping pattern is also seen in several high-resolution Doppler time-series from individual filament threads by \citet{lin04}, as well as in the Hinode/SOT observations by \citet{ning}, who reported a maximum number of 8 periods before the oscillations disappeared. In the context of the kink MHD mode interpretation, several mechanisms have been proposed to explain this quick damping \citep[see][]{oliver}. \citet{solernonad} studied the damping by nonadiabatic effects (radiative losses and thermal conduction) in a fully ionized, homogeneous, cylindrical filament thread embedded in a homogeneous corona, and obtained a kink mode damping time larger than $10^5$ periods for typical prominence conditions, meaning than nonadiabatic effects cannot explain the observed damping of transverse oscillations. Subsequently, \citet{arregui} considered a transverse inhomogeneous transitional layer between the fully ionized filament thread and the corona, and investigated the kink mode damping by resonant absorption in the Alfv\'en continuum. They obtained a damping time of approximately 3 periods for typical wavelengths of prominence oscillations. Later on, \citet[hereafter Paper~I]{solerslow} complemented the work by \citet{arregui} by also considering the damping by resonant absorption in the slow continuum, but concluded that this effect is not relevant and obtained similar results to those of \citet{arregui}. Finally, \citet[hereafter Paper~II]{solerneutrals} included the effect of partial ionization in a homogeneous filament thread model. These authors found that ion-neutral collisions can efficiently damp the kink mode for short wavelengths, although they are not efficient enough in the range of typically observed wavelengths.

On the basis of these previous studies, resonant absorption seems to be the most efficient damping mechanism for the kink mode and the only one that can produce the observed damping times. On the other hand, the effect of partial ionization could be also relevant, at least for short wavelengths. The aim of the present work is to take both mechanisms into account and to assess their combined effect on the kink mode damping. To our knowledge, this is the first time that the phenomenon of resonant absorption is studied in a partially ionized plasma. The filament thread model assumed here is similar to that of Paper~I and is composed of a homogeneous and straight cylinder with prominence-like conditions embedded in an unbounded corona, with a transverse inhomogeneous transitional layer between both media. We consider the prominence material to be partially ionized. As in Paper~II, we use the linearized MHD equations for a partially ionized, one-fluid plasma derived by \citet{forteza}. Since we focus our investigation on the kink mode, we adopt the $\beta = 0$ approximation, where $\beta$ is the ratio of the gas pressure to the magnetic pressure. This approximation removes longitudinal, slowlike modes but transverse modes remain correctly described. In the next Sections, we investigate the kink mode damping by means of both analytical and numerical computations, and assess the efficiency of the considered damping mechanisms.

This paper is organized as follows: Section~\ref{sec:math} contains a description of the model configuration and the basic equations. The results are presented and discussed in Section~\ref{sec:results}. Finally, our conclusions are given in Section~\ref{sec:conclusion}.

\section{MODEL AND METHOD}
\label{sec:math}

\subsection{Equilibrium Properties}	

We model a filament thread as an infinite and straight cylindrical magnetic flux tube of radius $a$ surrounded by an unbounded coronal environment. Cylindrical coordinates are used, namely $r$, $\varphi$, and $z$, for the radial, azimuthal, and longitudinal directions, respectively. In the following expressions, a subscript 0 indicates equilibrium quantities while we use subscripts $f$ and $c$ to explicitly denote filament and coronal quantities, respectively. The magnetic field is homogeneous and orientated along the cylinder axis, ${\mathit \bf{B}}_0 = B_0 \hat{e}_z$, with $B_0 = 5$~G constant everywhere. We adopt the one-fluid approximation and consider a hydrogen plasma composed of ions (protons), electrons, and neutral atoms. Since we are interested in kink MHD waves supported by the filament thread body, we also assume the $\beta = 0$ approximation, which neglects gas pressure effects and removes longitudinal, slowlike modes. With such assumptions \citep[see details in][and Paper II]{forteza}, the plasma properties are characterized by two quantities: the fluid density, $\rho_0$, and the ionization fraction, $\mut_0$, which gives us information about the plasma degree of ionization. The allowed values of $\mut_0$ range between $\mut_0 = 0.5$ for a fully ionized plasma and $\mut_0 = 1$ for a neutral plasma. 

The density profile assumed here only depends on the radial direction and is the same as in Paper I \citep[which was adopted after][]{rudermanroberts}, namely 
\begin{equation}
 \rho_{0}\left(r\right)=\left\{\begin{array}{clc}
 \rho_f,&{\rm if}&r\le a - l/2,   \\
 \rho_{\rm tr}\left(r\right),&{\rm if}&a-l/2<r<a+l/2,\\ 
 \rho_c,&{\rm if}&r\geq  a+l/2, \\
\end{array} \right. \label{eq:profilerho}
\end{equation}
with
\begin{equation}
 \rho_{\rm tr}\left(r\right)=\frac{\rho_f}{2}\left\{\left(1+\frac{\rho_c}{\rho_f}\right) - \left( 1-\frac{\rho_c}{\rho_f}\right)\sin \left[\frac{\pi}{l}\left( r-a\right)\right]\right\},
\end{equation}
where we take $\rho_f=5\times10^{-11}~{\rm kg}~{\rm m}^{-3}$ and $\rho_c=2.5\times 10^{-13}~{\rm kg}~{\rm m}^{-3}$, the density contrast between the filament and coronal plasma being $\rho_f/\rho_c=200$. In these expressions, $a$ is the thread mean radius and $l$ is the transitional layer width. The limit cases $l/a=0$ and $l/a=2$ correspond to a homogeneous thread without transitional layer and a fully inhomogeneous thread, respectively. We also take the same functional dependence for the ionization fraction profile,
 \begin{equation}
 \mut_{0}\left(r\right)=\left\{\begin{array}{clc}
 \mut_f,&{\rm if}&r\le a - l/2,   \\
 \mut_{\rm tr}\left(r\right),&{\rm if}&a-l/2<r<a+l/2,\\ 
 \mut_c,&{\rm if}&r\geq  a+l/2, \\
\end{array} \right. \label{eq:profilemu}
\end{equation}
with
\begin{equation}
 \mut_{\rm tr}\left(r\right)=\frac{\mut_f}{2}\left\{\left(1+\frac{\mut_c}{\mut_f}\right) - \left( 1-\frac{\mut_c}{\mut_f}\right)\sin \left[\frac{\pi}{l}\left( r-a\right)\right]\right\},
\end{equation}
where the filament ionization fraction, $\mut_f$, is considered a free parameter and the corona is assumed to be fully ionized, so $\mut_c = 0.5$.

\subsection{Basic Equations}

\begin{figure}[!htb]
\centering
\epsscale{0.99}
\plotone{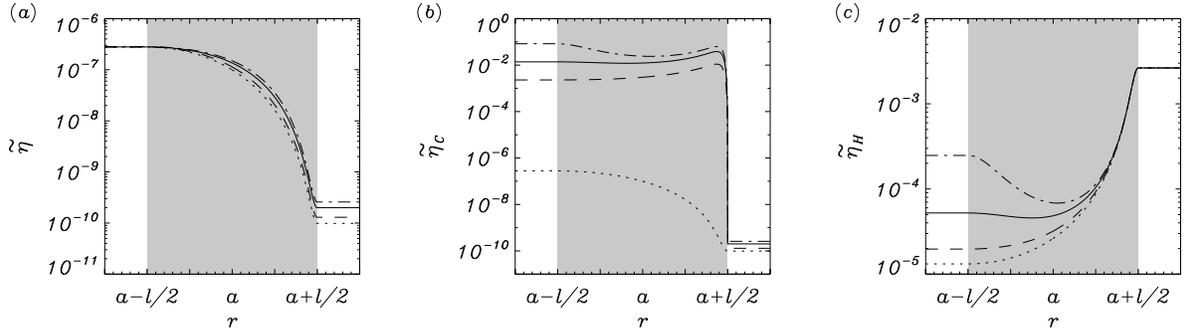}
\caption{Radial profiles of ($a$) $\etat$, ($b$) $\etact$, and ($c$) $\etaht$ considered in this work. The transitional layer (shaded zone) is enhanced in order to see the change from filament to coronal values. The linestyles represent different ionization degrees: $\tilde{\mu}_f = 0.5$ (dotted line),  $\tilde{\mu}_f = 0.6$ (dashed line), $\tilde{\mu}_f = 0.8$ (solid line), and $\tilde{\mu}_f = 0.95$ (dash-dotted line).\label{fig:prof}}
\end{figure}

We consider the basic MHD equations for a partially ionized plasma in the one-fluid approach derived by \citet{forteza} \citep[see also][]{bragi}. Since we are dealing with small-amplitude perturbations, we restrict ourselves to linear perturbations. After removing gas pressure terms by setting $\beta = 0$, the relevant equations for our investigation are the momentum and the induction equations, whose linearized version in MKS units are,
\begin{equation}
 \rho_0  \frac{\pd {\mathit {\bf v}}_1}{\pd t}  =  \frac{1}{\mu} \left[ \left( \nabla \times {\mathit {\bf B}}_1  \right) \times {\mathit {\bf B}}_0 \right], \label{eq:momentum}
\end{equation}
\begin{equation}
  \frac{\pd {\mathit {\bf B}}_1}{\pd t} = \nabla \times \left( {\mathit {\bf v}}_1 \times {\mathit {\bf B}}_0\right) - \nabla \times \left( \eta \nabla \times {\mathit {\bf B}}_1 \right) +  \nabla \times \left\{  \eta_{\rm A} \left[ \left( \nabla \times {\mathit {\bf B}}_1  \right) \times {\mathit {\bf B}}_0 \right] \times {\mathit {\bf B}}_0  \right\} - \nabla \times \left[  \eta_{\rm H} \left( \nabla \times {\mathit {\bf B}}_1  \right) \times {\mathit {\bf B}}_0 \right] , \label{eq:induction}
\end{equation}
along with the condition $\nabla \cdot {\mathit {\bf B}}_1 = 0$. In these equations ${\mathit {\bf B}}_1 = \left( B_r , B_\varphi, B_z \right)$, ${\mathit {\bf v}}_1 = \left( v_r , v_\varphi, v_z \right)$ are the components of the magnetic field and velocity perturbations, respectively, while  $\mu = 4 \pi \times 10^{-7}$~N~A$^{-2}$ is the vacuum magnetic permeability. Quantities $\eta$, $\eta_{\rm A}$, and $\etah$ in Equation~(\ref{eq:induction}) are the coefficients of ohmic, ambipolar, and Hall's magnetic diffusion. Note that these nonideal terms of the induction equation appear due to collisions between the different plasma species. In particular, ohmic diffusion is mainly governed by electron-ion collisions and ambipolar diffusion is mostly caused by ion-neutral collisions. Hall's effect is enhanced by ion-neutral collisions since they tend to decouple ions from the magnetic field while electrons remain able to drift with the magnetic field \citep{pandey}. The ambipolar diffusivity is commonly expressed in terms of the Cowling's coefficient, $\etac$, as follows,
\begin{equation}
 \eta_{\rm A} = \frac{\etac - \eta}{B_0^2}.
\end{equation}
As explained in Paper~II, $\eta$ and $\etac$ correspond to the magnetic diffusivities longitudinal and perpendicular to magnetic field lines, respectively. Due to the presence of neutrals, $\etac \gg \eta$, meaning that perpendicular magnetic diffusion is much more efficient than longitudinal magnetic diffusion in a partially ionized plasma. In a fully ionized plasma $\etac = \eta$, so magnetic diffusion is then isotropic. Since $\eta$, $\etac$, and $\etah$ are functions of the plasma physical conditions \citep[see, e.g.,][Paper II for expressions]{bragi,khoda04,leake05,leake06,pandey}, their values in our equilibrium depend on the radial direction. It is convenient for our following analysis to write ohmic, Cowling's, and Hall's diffusivities in a dimensionless form,
\begin{equation}
 \etat = \frac{\eta}{\vap a}, \qquad  \etact = \frac{\etac}{\vap a}, \qquad \etaht = \frac{\etah B_0}{\vap a}, \label{eq:etasadimen}
\end{equation}
where the tilde denotes the dimensionless quantity and $\vap = B_0 / \sqrt{ \mu_0 \rho_f}$ is the filament Alfv\'en speed. Note that $\etat^{-1}$ is usually called the magnetic Reynolds number. Figure~\ref{fig:prof} displays the radial profiles of $\etat$, $\etact$, and $\etaht$ according to the equilibrium properties. The Figure focuses in the transitional layer, where the dimensionless coefficients vary by several orders of magnitude from internal to external values. 

Since $\varphi$ and $z$ are ignorable coordinates, perturbations are put proportional to $\exp \left( i m \varphi + i k_z z - i \omega t \right)$, where $\omega$ is the oscillatory frequency, and $m$ and $k_z$ are the azimuthal and longitudinal wavenumbers, respectively. Then, Equations~(\ref{eq:momentum})--(\ref{eq:induction}) become,
\begin{equation}
 \omega v_r = - \frac{\va^2}{B_0} \left( k_z B_r + i B'_z \right), \label{eq:vr}
\end{equation}
\begin{equation}
 \omega v_\varphi = \frac{\va^2}{B_0} \left( \frac{m}{r} B_z - k_z B_\varphi \right), \label{eq:vphi}
\end{equation}
\begin{equation}
 \omega v_z = 0, \label{eq:vz} 
\end{equation}
\begin{equation}
 \omega B_r = - B_0 k_z v_r + \eta \left( \frac{m}{r} B'_\varphi + \frac{m}{r^2} B_\varphi - i  \frac{m^2}{r^2} B_r  \right) + \etac \left( k_z B'_z - i k_z^2 B_r \right) + \etah B_0 \left( i k_z \frac{m}{r} B_z - i k_z^2 B_\varphi \right), \label{eq:br}
\end{equation}
\begin{eqnarray}
 \omega B_\varphi &=& -B_0 k_z v_\varphi + \eta \left( i B''_\varphi + i \frac{1}{r} B'_\varphi - i \frac{1}{r^2} B_\varphi + \frac{m}{r} B'_r - \frac{m}{r^2} B_r \right) + \eta' \left( i B'_\varphi + i \frac{1}{r} B_\varphi + \frac{m}{r} B_r \right) \nonumber \\ &&+ \etac \left( i k_z \frac{m}{r} B_z - i k_z^2 B_\varphi \right) + \etah B_0 \left( i k_z^2 B_r - k_z B'_z \right), \label{eq:bphi}
\end{eqnarray}
\begin{eqnarray}
 \omega B_z &=& - B_0 \left( i v'_r + i \frac{1}{r} v_r - \frac{m}{r} v_\varphi \right) + \etac \left( i B''_z + i \frac{1}{r} B'_z - i \frac{m^2}{r^2} B_z + k_z B'_r + k_z \frac{1}{r} B_r + i k_z \frac{m}{r} B_\varphi  \right)  \nonumber \\ &&+ \etac' \left( k_z B_r + i B'_z \right) + \etah B_0 \left( k_z B'_\varphi + k_z \frac{1}{r} B_\varphi - k_z \frac{m}{r} B_r \right) + \etah' B_0 \left( k_z B_\varphi - \frac{m}{r} B_z \right), \label{eq:bz}
\end{eqnarray}
where the prime denotes derivative with respect to $r$. From Equation~(\ref{eq:vz}) we get $v_z=0$, so no longitudinal displacements are allowed in the $\beta = 0$ approximation. These equations form an eigenvalue problem, with $\omega$ the eigenvalue and $\left( v_r, v_\varphi, B_r, B_\varphi, B_z \right)$ the eigenvector. In the present work, we study the time damping of kink waves, so we assume $m=1$ and take a real and positive $k_z$. A complex frequency, $\omega = \omega_{\rm R} + i \omega_{\rm I}$, is therefore expected, damped solutions corresponding to $\omega_{\rm I} < 0$. The period, $P$, and damping time, $\td$, are related to the real and imaginary parts of the frequency as follows,
\begin{equation}
 P = \frac{2 \pi}{\omega_{\rm R}}, \qquad \tau_{\rm D} = \frac{1}{\left|\omega_{\rm I}\right|}.
\end{equation}

\subsubsection{Importance of Hall's Term}
\label{sec:dimensional}

In Paper II, we showed that ohmic diffusion can be neglected in front of Cowling's diffusion for large $k_z a$, while ohmic diffusion dominates for small values of $k_z a$. In the present work, we have also included Hall's diffusion. Here, we compare the importance of Hall's term with the other nonideal terms in the induction equation. To achieve this, we first have to assess the typical length-scales of the dissipative terms of the induction equation. We note that longitudinal derivatives only appear in the terms with $\etact$ and $\etaht$ of Equations~(\ref{eq:br})--(\ref{eq:bz}), while the terms with $\etat$ only contain radial and azimuthal derivatives. Therefore, the typical length-scale of both Cowling's and Hall's terms is the longitudinal wavelength, $\lambda_z$, which can be expressed in terms of the longitudinal wavenumber, $\lambda_z = 2 \pi k_z^{-1}$. On the other hand, the typical length-scale of the ohmic term is the filament thread mean radius, $a$. By taking into account these relevant length-scales, we define the dimensionless number $\mathcal{H}$ as the magnitude of Hall's term with respect to that of the ohmic term,
\begin{equation}
 \mathcal{H} = \frac{\left| \nabla \times \left[  \eta_{\rm H} \left( \nabla \times {\mathit {\bf B}}_1  \right) \times {\mathit {\bf B}}_0 \right]  \right|}{\left|  \nabla \times \left( \eta \nabla \times {\mathit {\bf B}}_1 \right)  \right|} \sim \left( \frac{k_z a}{2 \pi} \right)^2  \frac{\etaht}{\etat}. \label{eq:H}
\end{equation}
From Equation~(\ref{eq:H}), one can compute the value of $k_z a$ for which Hall's term becomes more important than the Ohm's term by setting $ \mathcal{H} \approx 1$. So, one obtains,
\begin{equation}
 \left( k_z a \right)_{\rm H} \approx 2 \pi \sqrt{\frac{\etat}{\etaht}}.
\end{equation}
We can compare this transitional wavenumber with that obtained in Section 2.3. of Paper II that determines when Cowling's diffusion begins to dominate over ohmic diffusion,
\begin{equation}
 \left(  k_z a \right)_{\rm C} \approx 2 \pi \sqrt{\frac{\etat}{\etact}}. \label{eq:transC}
\end{equation}
Since in our equilibrium $\etact > \etaht$,  one gets $\left(  k_z a \right)_{\rm C} <  \left( k_z a \right)_{\rm H}$, meaning than Cowling's diffusion becomes dominant over ohmic diffusion for smaller wavenumbers than those needed for Hall's term to become relevant. Next, we have to know whether Cowling's diffusion or Hall's diffusion dominate for $k_z a$ beyond both transitional values. To do so, we define the dimensionless number $\mathcal{H}_{\rm C}$ as the magnitude of Hall's term with respect to that of the ambipolar term,
\begin{equation}
 \mathcal{H_{\rm C}} = \frac{\left| \nabla \times \left[  \eta_{\rm H} \left( \nabla \times {\mathit {\bf B}}_1  \right) \times {\mathit {\bf B}}_0 \right]  \right|}{\left| \nabla \times \left\{  \eta_{\rm A} \left[ \left( \nabla \times {\mathit {\bf B}}_1  \right) \times {\mathit {\bf B}}_0 \right] \times {\mathit {\bf B}}_0  \right\}  \right|} \sim \frac{\etaht}{\etact}.
\end{equation}
Considering again that $\etact > \etaht$, the condition $\mathcal{H_{\rm C}} <1$ is always satisfied, therefore Cowling's diffusion is always more important than Hall's diffusion. By means of this simple dimensional analysis, we expect the presence of Hall's term to have a minor effect on the results.

\subsection{Analytical Dispersion Relation}

Some analytical progress can be performed when no transitional layer is present, i.e., $l/a = 0$. We showed in Paper~II that it is possible to give an analytical dispersion relation when the terms with $\eta$ and $\etah$ are dropped from the induction equation. In such a situation, the induction equation (Equation~(\ref{eq:induction})) can be written in a compact form as follows,
\begin{equation}
  \frac{\pd {\mathit {\bf B}}_1}{\pd t} = \frac{\Gamma^2_{\rm A} }{\va^2} \nabla \times \left( {\mathit {\bf v}}_1 \times {\mathit {\bf B}}_0\right), \label{eq:induction2}
\end{equation}
where $\Gamma^2_{\rm A} \equiv \va^2 - i \omega \etac$ is the modified Alfv\'en speed squared \citep{fortezanoad}. Next, we follow the standard procedure \citep[see, e.g.,][]{edwinroberts,goossens} and obtain the dispersion relation of trapped waves by imposing that  the radial displacement, $\xi_r = i v_r / \omega$, and the total pressure perturbation, $p_{\rm T} = B_0 B_z / \mu$, are both continuous at the thread boundary, $r = a$, and that perturbations must vanish at infinity. So,  the dispersion relation governing transverse oscillations is $\mathcal{D}_m \left( \omega, k_z \right) = 0$, with
\begin{equation}
\mathcal{D}_m \left( \omega, k_z \right) =  \frac{n_c}{\rho_c \left( \omega^2 -  k_z^2 \Gamma_{{\rm A}c}^2 \right)} \frac{K'_m \left( n_c a \right)}{K_m \left( n_c a \right)} - \frac{m_f}{\rho_f \left( \omega^2 - k_z^2 \Gamma_{{\rm A}f}^2 \right)} \frac{J'_m \left( m_f a \right)}{J_m \left( m_f a \right)}, \label{eq:dispernocapa}
\end{equation}
where $J_m$ and $K_m$ are the Bessel function and the modified Bessel function of the first kind \citep{abramowitz}, respectively, and the quantities $m_f$ and $n_c$ are given by
\begin{equation}
 m_f^2 = \frac{\left( \omega^2 - k_z^2 \Gamma_{{\rm A}f}^2  \right)}{\Gamma_{{\rm A}f}^2}, \qquad n_c^2 = \frac{\left( k_z^2 \Gamma_{{\rm A}c}^2 - \omega^2  \right)}{\Gamma_{{\rm A}c}^2}.
\end{equation}
Note that Equation~(\ref{eq:dispernocapa}) is valid for any value of $m$.

Next, our aim is to extend this analytical analysis to the case $l/a \neq 0$. When an inhomogeneous transitional layer is present in the equilibrium, the kink mode is resonantly coupled to Alfv\'en continuum modes. The radial position where the Alfv\'en resonance takes place, $r_{\rm A}$, can be computed by setting the kink mode frequency, $\omega_k$, equal to the local Alfv\'en frequency, $\omega_{\rm A} = \va k_z$. The expression of $r_{\rm A}$ corresponding to our density profile was obtained in Paper~I (Equation~(9))\footnote{Note that there is a typographical error in Equation~(9) of Paper~I since the term $\rho_e$ (which corresponds to our $\rho_c$) should be $-\rho_e$.},
\begin{equation}
  r_{\rm A} = a + \frac{l}{\pi} \arcsin \left[ \frac{\rho_f + \rho_c}{\rho_f - \rho_c} - \frac{2 \vap^2 k_z^2}{\omega_k^2} \frac{\rho_f }{\left( \rho_f - \rho_c \right)}  \right]. \label{eq:resonantpoint}
\end{equation}
The ideal MHD equations are singular at $r = r_{\rm A}$. This singularity is removed if dissipative effects, such as magnetic diffusion or viscosity, are considered in a region around the resonance point, i.e., the dissipative layer. A method to obtain an analytical dispersion relation in the presence of an inhomogeneous transitional layer is to combine the jump conditions at the resonance point with the so-called thin boundary (TB) approximation, which was first used by \citet{hollwegyang}. The jump conditions were first derived by \citet{SGH91} and \citet{goossens95} for the driven problem, and later by \citet{tirry} for the eigenvalue problem. They have been used in a number of papers in the context of MHD waves in the solar atmosphere \citep[e.g.,][Paper~I among other works]{sakuraib, goossens92,keppens,stenuit,andries2000,tom}. An important result for the present investigation was obtained by \citet{goossens95}, who proved that the jump conditions derived by \citet{SGH91} in ideal MHD remain valid in dissipative MHD. This allows us to apply the jump conditions derived by \citet{SGH91} to our case. The assumptions behind the TB approximation and its applications have been recently reviewed by \citet{goossensIAU}. In short, the main assumption of the TB approximation is that there is a region around the dissipative layer where both ideal and dissipative MHD applies. In our equilibrium, we can assume that the thickness of the dissipative layer, namely $\delta_{\rm A}$, roughly coincides with the width of the inhomogeneous transitional layer. This condition is approximately verified for thin layers, i.e., $l/a \ll 1$. So, we can simply connect analitycally the perturbations from the homogeneous part of the tube to those of the external medium by means of the jump conditions and avoid the numerical integration of the dissipative MHD equations across the inhomogeneous transitional layer.

The jump conditions for the radial displacement and the total pressure perturbation provided by \citet{SGH91} in the case of a straight magnetic field are,
\begin{equation}
 \left[ \left[ \xi_r \right]\right] = - i \pi \frac{m^2 / r_{\rm A}^2}{\left| \rho_0 \Delta \right|_{r_{\rm A}}} p_{\rm T}, \qquad  \left[ \left[ p_{\rm T} \right] \right]= 0, \quad \textrm{at} \quad r= r_{\rm A}, \label{eq:boundarycond}
\end{equation}
where $\left[ \left[ X \right]\right] = X_c - X_f$ stands for the jump of the quantity $X$, and $\Delta = \frac{\rm d}{{\rm d}r} \left( \omega^2 - \omega_{\rm A}^2 \right)$. By considering that in our model the magnetic field is straight and constant so that the variations of the local Alfv\'en frequency are only due to the variation of the equilibrium density, we can write $\left| \rho_0 \Delta \right|_{r_{\rm A}} = \omega_{\rm A}^2 \left| \partial_r \rho_0  \right|_{r_{\rm A}}$. In addition, from the resonance condition we have $\omega_{\rm A}^2 = \omega_k^2$. Applying the jump conditions~(\ref{eq:boundarycond}), we arrive at the dispersion relation in the TB approximation,
\begin{equation}
\mathcal{D}_m \left( \omega, k_z \right) =  -i \pi \frac{m^2 / r_{\rm A}^2}{\omega_k^2 \left| \partial_r \rho_0  \right|_{r_{\rm A}}}, \label{eq:dispercapa}
\end{equation}
with $\mathcal{D}_m \left( \omega, k_z \right)$ defined in Equation~(\ref{eq:dispernocapa}). Note that in order to solve Equation~(\ref{eq:dispercapa}) we need the value of the kink frequency, $\omega_k$. For this reason, we use a two-step procedure. First, we solve the dispersion relation for the case $l/a = 0$ (Equation~(\ref{eq:dispernocapa})) and obtain $\omega_k$. Next, we assume that the real part of the frequency is approximately the same when the inhomogeneous transitional layer is included, allowing us to determine $r_{\rm A}$ from Equation~(\ref{eq:resonantpoint}) and therefore $ \left| \partial_r \rho_0  \right|_{r_{\rm A}}$. Finally, we solve the complete dispersion relation (Equation~(\ref{eq:dispercapa})) with these parameters.

\subsubsection{Expressions in the Thin Tube Limit}
\label{sec:tt}

Equation~(\ref{eq:dispercapa}) is a transcendental equation that has to be solved numerically. As in Paper~I, it is possible to go further analytically by considering the thin tube (TT) approximation, i.e., $k_z a \ll 1$. The combination of both the TB and TT approximations has been done in several works \citep[e.g.,][]{goossens92, rudermanroberts, goossens02, goossens}. In the TT limit, we approximate the kink mode frequency as 
\begin{equation}
\omega_k \approx \sqrt{\frac{2}{ 1 + \rho_c / \rho_f }} \vap k_z.
\end{equation}
We put this expression in Equation~(\ref{eq:resonantpoint}) and obtain that $r_{\rm A} \approx a$. Next we perform a first order, asymptotic expansion for small arguments of the Bessel functions of Equation~(\ref{eq:dispercapa}). The dispersion relation then becomes,
\begin{equation}
\rho_f \left( \omega^2 -  k_z^2 \Gamma_{{\rm A}f}^2 \right) + \rho_c \left( \omega^2 -  k_z^2 \Gamma_{{\rm A}c}^2 \right) =  i \pi \left( \frac{m}{a} \right) \frac{\rho_f \rho_c}{\left| \partial_r \rho_0  \right|_{r_{\rm A}}} \frac{\left( \omega^2 -  k_z^2 \Gamma_{{\rm A}f}^2 \right) \left( \omega^2 -  k_z^2 \Gamma_{{\rm A}c}^2 \right)}{\omega_k^2 }. \label{eq:disperTT}
\end{equation}
We now write the frequency as $\omega = \omega_k + i \omega_{\rm I}$, and the modified Alfv\'en speed squared is approximated by $\Gamma_{\rm A}^2 \approx \va^2 - i \omega_k \etac$. We insert these expressions in Equation~(\ref{eq:disperTT}) and neglect terms with $\omega_{\rm I}^2$ and $\omega_{\rm I} k_z^2$. It is straight-forward to obtain an expression for the ratio $\omega_{\rm I} / \omega_k$,
\begin{eqnarray}
 \frac{\omega_{\rm I}}{\omega_k} = &-& \frac{\pi}{2} \left( \frac{m}{a} \right) \frac{\rho_f \rho_c}{\left(\rho_f + \rho_c \right)} \frac{1}{ \left| \partial_r \rho_0  \right|_{r_{\rm A}}} \left[ \frac{1}{4} \frac{\left(\rho_f - \rho_c \right)^2}{\rho_f \rho_c} + \left( \frac{1}{\vap^2} + \frac{1}{\vac^2}  \right) \frac{{\etac}_f {\etac}_c}{2} k_z^2  \right] \nonumber \\ &-& \frac{1}{2} \frac{\left(\rho_f {\etac}_f + \rho_c {\etac}_c \right)k_z}{\sqrt{\left(\rho_f + \rho_c \right) \left(\rho_f \vap^2+ \rho_c \vac^2\right)}}. \label{eq:wiwk}
\end{eqnarray}
The first term in Equation~(\ref{eq:wiwk}) owes its existence to the factor in the dispersion relation related to the TB approximation and represents the contribution of resonant absorption. For long wavelengths, the factor $\left( 1/\vap^2 + 1/\vac^2 \right){\etac}_f {\etac}_c k_z^2 / 2$ can be neglected, so the term related to the resonant damping is independent of the value of Cowling's diffusivity and, therefore, of the ionization degree. Then, one can see that in the TT case the term of Equation~(\ref{eq:wiwk}) due to the resonant damping takes the same form as in a fully ionized plasma, which was previously obtained by, e.g., \citet[Equation~(77)]{goossens92} and \citet[Equation~(56)]{rudermanroberts}. On the other hand, the second term in Equation~(\ref{eq:wiwk}) is related to the damping by Cowling's diffusion and is also present in the case $l/a = 0$. This term is proportional to $k_z$, so we also expect it to be of a minor influence in the TT regime.

Next, we take into account that for the present sinusoidal density profile, $ \left| \partial_r \rho_0  \right|_{r_{\rm A}} \approx \pi \left( \rho_f - \rho_c \right) / 2 l$. We insert this expression in Equation~(\ref{eq:wiwk}) and then use it to give a relation for the ratio of the damping time to the period,
\begin{equation}
 \frac{\td}{P} = \frac{2}{\pi} \left[ m \left( \frac{l}{a} \right) \left(  \frac{\rho_f - \rho_c}{\rho_f + \rho_c } \right) + \frac{2 \left(\rho_f \etactf + \rho_c \etactc \right)k_z a}{\sqrt{2 \rho_f \left(\rho_f + \rho_c \right) }} \right]^{-1}, \label{eq:tdp}
\end{equation}
where we have used Equation~(\ref{eq:etasadimen}) to rewrite Cowling's diffusivities in their dimensionless form. To perform a simple application, we compute $\tdp$ from Equation~(\ref{eq:tdp}) in the case $m=1$, $k_z a = 10^{-2}$, and $l/a = 0.2$, resulting in $\tdp \approx 3.18$ for a fully ionized tread ($\mut_f = 0.5$), and  $\tdp \approx 3.16$ for an almost neutral thread ($\mut_f = 0.95$). We note that the obtained damping times are consistent with the observations. Furthermore, the ratio $\tdp$ depends very slightly on the ionization degree, suggesting that resonant absorption dominates over Cowling's diffusion. To check this last statement, we compute the ratio of the two terms on the right-hand side of Equation~(\ref{eq:tdp}), which allows us to compare the damping times exclusively due to resonant absorption, $\left( \td \right)_{\rm RA}$, and Cowling's diffusion, $\left( \td \right)_{\rm C}$,
\begin{equation}
 \frac{\left( \td \right)_{\rm RA}}{\left( \td \right)_{\rm C}} = \sqrt{\frac{2 \left(\rho_f + \rho_c \right)}{\rho_f}} \left( \frac{\rho_f \etactf + \rho_c \etactc}{\rho_f - \rho_c} \right) \frac{k_z a}{m \left( l/a \right)}.
\end{equation}
This last expression can be further simplified by considering that in filament threads $\rho_f \gg \rho_c$ and $\etactf \gg \etactc$, so that
\begin{equation}
 \frac{\left( \td \right)_{\rm RA}}{\left( \td \right)_{\rm C}} \approx \sqrt{2} \etactf \frac{k_z a}{m \left( l/a \right)}. \label{eq:ratiotaus}
\end{equation}
Thus, we see that the efficiency of Cowling's diffusion with respect to that of resonant absorption increases with $k_z a$ and $\mut_f$ (through $\etactf$). Considering the same parameters as before, one obtains $\left( \td \right)_{\rm RA} / \left( \td \right)_{\rm C} \approx 2 \times 10^{-8}$ for $\mut_f = 0.5$, and $\left( \td \right)_{\rm RA} / \left( \td \right)_{\rm C} \approx 6 \times 10^{-3}$ for $\mut_f = 0.95$, meaning that resonant absorption is much more efficient than Cowling's diffusion. From Equation~(\ref{eq:ratiotaus}) it is also possible to give an estimation of the wavenumber for which Cowling's diffusion becomes more important than resonant absorption by setting $\left( \td \right)_{\rm RA} / \left( \td \right)_{\rm C} \approx 1$. So, one gets,
\begin{equation}
 k_z a \approx \frac{m \left( l/a \right)}{\sqrt{2} \etactf}. \label{eq:kzCRA}
\end{equation}
Considering again the same parameters, Equation~(\ref{eq:kzCRA}) gives $k_z a \approx 5 \times 10^5$ for $\mut_f = 0.5$, and $k_z a \approx 1.7$ for $\mut_f = 0.95$. One has to bear in mind that Equation~(\ref{eq:ratiotaus}) is valid only for $k_z a \ll 1$, so we expect resonant absorption to be the dominant damping mechanism in the TT regime even for an almost neutral filament plasma. We will verify the analytical estimations of the present Section by means of numerical computations.

\subsection{Numerical Computations}

To numerically solve the full eigenvalue problem (Equations~(\ref{eq:vr})--(\ref{eq:bz})) we use the PDE2D code based on finite elements \citep{sewell}. We follow the same procedure as in Papers~I and II. The integration of Equations~(\ref{eq:vr})--(\ref{eq:bz}) is performed from the cylinder axis, $r=0$, to the edge of the numerical domain, $r=r_{\rm max}$, where all perturbations vanish since the evanescent condition is imposed in the coronal medium, i.e., we restrict ourselves to trapped modes. The boundary conditions at $r=0$ are imposed by symmetry arguments. To obtain a good convergence of the solution and to avoid numerical errors, we need to locate the edge of the numerical domain far enough from the filament thread mean radius to satisfy the evanescent condition. We have considered $r_{\rm max} = 100 a$. We use a nonuniform grid with a large density of grid points within the inhomogeneous transitional layer in order to correctly describe the small spatial scales that develop due to the Alfv\'en resonance.

\section{RESULTS}
\label{sec:results}

We focus our study on the ratio of the damping time to the period, $\tdp$, which is the quantity that informs us about the efficiency of the kink mode damping. In the following sections, we compute $\tdp$ as a function of the dimensionless longitudinal wavenumber, $k_z a$. According to the observed wavelengths \citep{oliverballester} and thread widths \citep{lin04}, the relevant range of $k_z a$ of filament oscillations corresponds to $10^{-3} < k_z a < 10^{-1}$. So, the results within this range of $k_z a$ will deserve special attention.

\subsection{Filament Thread without Transitional Layer}
\label{sec:nocapa}

\begin{figure*}[!htp]
\centering
\epsscale{0.49}
\plotone{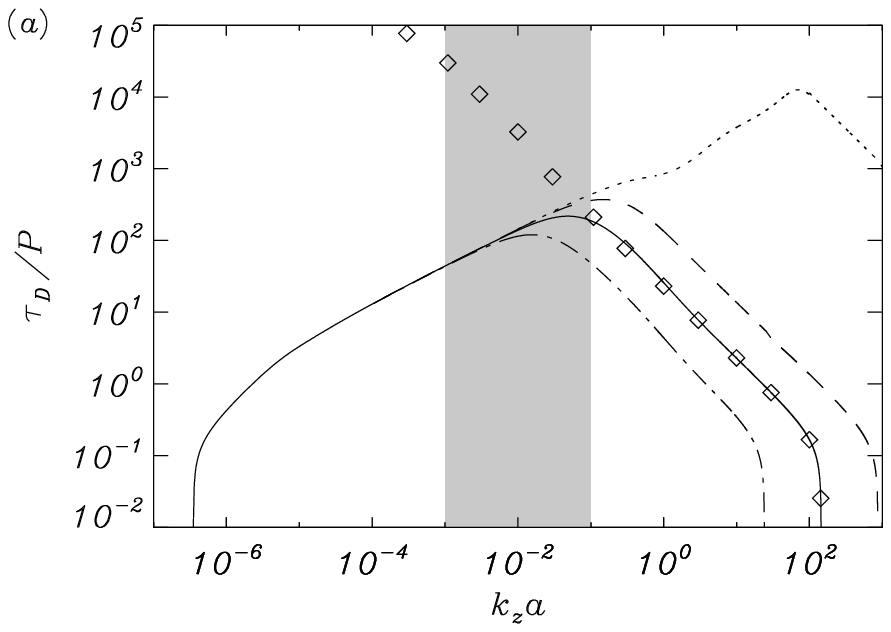}
\plotone{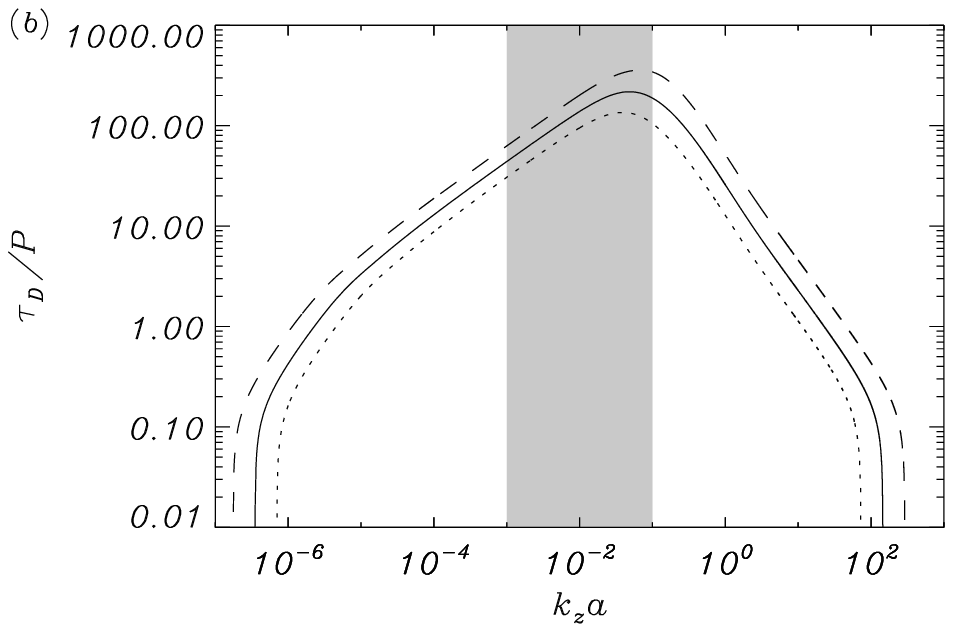}
\caption{Ratio of the damping time to the period of the kink mode as a function of $k_z a$ corresponding to a thread without transitional layer, i.e., $l/a=0$. ($a$) Results for $a = 100$~km considering different ionization degrees: $\tilde{\mu}_f = 0.5$ (dotted line),  $\tilde{\mu}_f = 0.6$ (dashed line), $\tilde{\mu}_f = 0.8$ (solid line), and $\tilde{\mu}_f = 0.95$ (dash-dotted line). Symbols are the approximate solution given by solving Equation~(\ref{eq:dispernocapa}) for $\tilde{\mu}_f = 0.8$. ($b$) Results for  $\tilde{\mu}_f = 0.8$ considering different thread widths: $a = 100$~km (solid line), $a = 50$~km (dotted line), and $a = 200$~km (dashed line). The shaded zone corresponds to the range of typically observed wavelengths of prominence oscillations.  \label{fig:nocapa}}
\end{figure*}

First, we start with the case of a homogeneous filament thread without transitional layer, i.e., $l/a = 0$. This is the configuration studied in Paper~II with the addition of Hall's term in the induction equation. Hence, we briefly discuss these results here and refer the reader to Paper~II for a more complete description. Figure~\ref{fig:nocapa}($a$) displays $\tdp$ versus $k_z a$ for different ionization degrees. In agreement with Paper~II, we obtain that $\tdp$ has a maximum which corresponds to the transition between the ohmic-dominated regime, which is almost independent of the ionization degree, to the region where Cowling's diffusion is more relevant and the ionization degree has a significant influence. The position of this transitional wavenumber is well approximated by Equation~(\ref{eq:transC}). As expected, the approximate solution obtained by solving Equation~(\ref{eq:dispernocapa}) for $\mut = 0.8$ (symbols in Figure~\ref{fig:nocapa}($a$)) agrees well with the numerical solution in the range of $k_z a$ where Cowling's diffusion dominates, while it significantly diverges from the numerical solution in the region where ohmic diffusion is relevant. Within the range of typically reported wavelengths of filament oscillations (shaded region), $\tdp$ is between 1 and 2 orders of magnitude larger than the observed values, which implies that neither ohmic nor Cowling's diffusion provide realistic damping times compatible with those observed. 

On the other hand, the kink mode exists as a propagating wave for $k_z a$ in the range $k_z^{c-} a < k_z a < k_z^{c+} a$, where $k_z^{c-}$ and $k_z^{c+}$ are the critical wavenumbers described by Equation~(38) of Paper~II. With the notation adopted in the present work, these critical wavenumbers can be expressed in dimensionless form as follows,
\begin{equation}
 k_z^{c-} a \approx \frac{\etat}{2} \left( k_\perp a \right)^2, \qquad k_z^{c+} a \approx \frac{2}{\etact},
\end{equation}
where $k_\perp$ is the perpendicular wavenumber to the magnetic field lines and contains the effect of the model geometry (see details in Paper~II). For our present cylindrical filament thread, this geometry-related factor can be written as $\left( k_\perp a \right)^2 \approx \left( k_r a \right)^2 + \left( k_\varphi a \right)^2$, where $k_r$ and $k_\varphi$ are the radial and azimuthal wavenumbers, respectively. We approximate $k_r^2$ by its expression in the ideal case,
\begin{equation}
 k_r^2 \approx \frac{\omega_k^2 - k_z^2 \vap^2}{\vap^2},
\end{equation}
which for $k_z a \ll 1$ and $\rho_f \gg \rho_c$ simplifies to $k_r^2 \approx k_z^2$. On the other hand, the azimuthal wavenumber can be expressed as $k_\varphi^2 \approx m^2/a^2$. In the long-wavelength case, the azimuthal wavenumber dominates over the radial wavenumber, so we get $\left( k_\perp a \right)^2 \approx m^2$. Next, we plot in Figure~\ref{fig:nocapa}($b$) the ratio of the damping time to the period as a function of $k_z a$ for $\mut_f = 0.8$ considering three different values of the filament thread radius. We see that the smaller the thread radius, the smaller $\tdp$ and so the more attenuated the kink wave. In addition, the critical wavenumbers are shifted when $a$ changes, the range of $k_z a$ for which the kink wave propagates being wider for thick threads than for thin threads. However, one must bear in mind that the observed width of filament threads is in the range $0.''2 - 0.''6$ \citep{lin04}, and therefore $a$ ranges from 75~km to 375~km, approximately. The critical wavenumbers are far from the relevant values of $k_z a$ for the observed thread widths, and so they should not affect the kink wave propagation in realistic filament threads.

\subsection{Filament Thread with an Inhomogeneous Transitional Layer}
\label{sec:capa}

\begin{figure*}[!p]
\centering
\epsscale{0.49}
\plotone{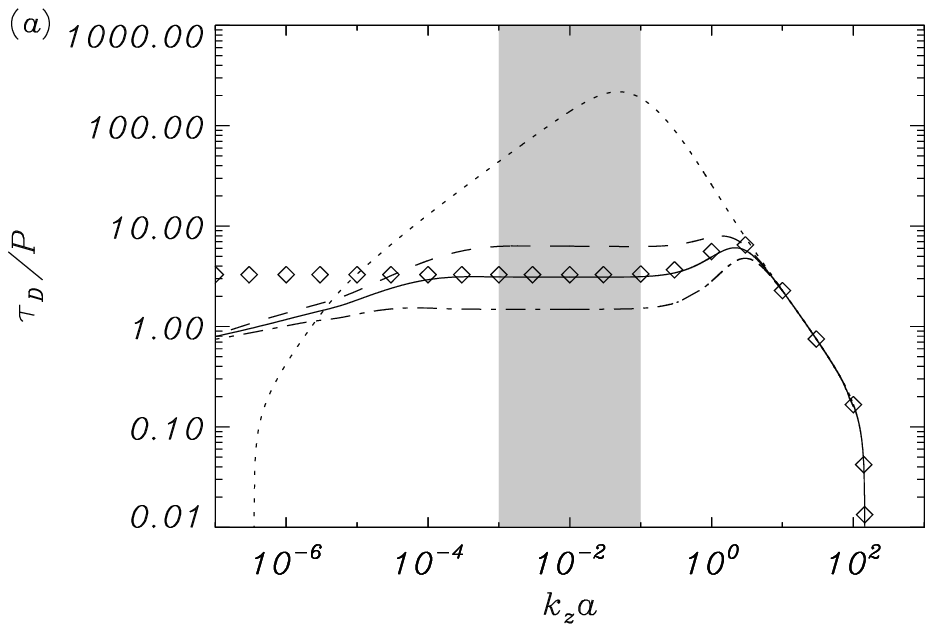}
\plotone{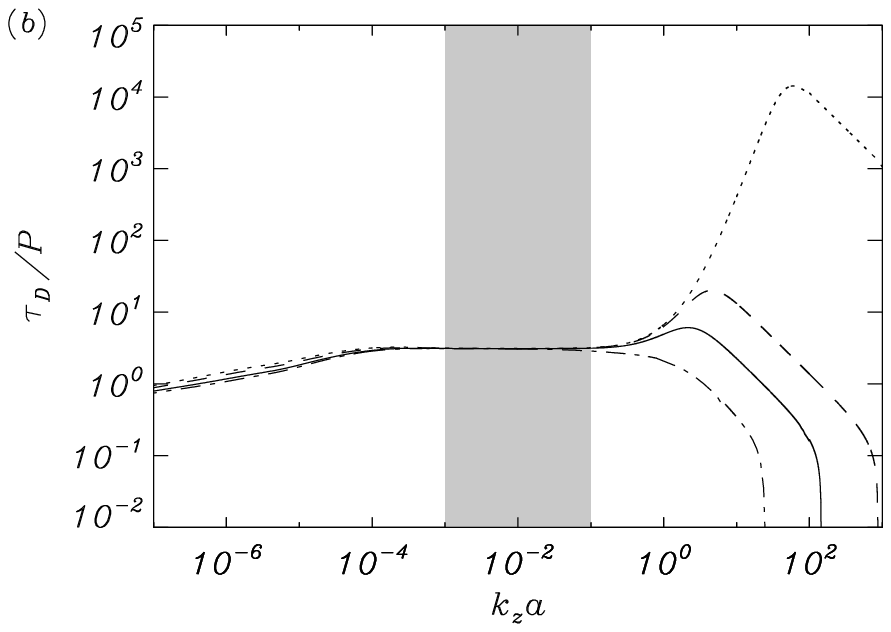}
\caption{Ratio of the damping time to the period of the kink mode as a function of $k_z a$ corresponding to a thread with an inhomogeneous transitional layer. ($a$) Results for $\mut_f = 0.8$ considering different transitional layer widths: $l/a = 0$ (dotted line),  $l/a = 0.1$ (dashed line), $l/a = 0.2$ (solid line), and $l/a  = 0.4$ (dash-dotted line). Symbols are the solution in the TB approximation given by solving Equation~(\ref{eq:dispercapa}) for $l/a = 0.2$. ($b$) Results for  $l/a = 0.2$ considering different ionization degrees: $\tilde{\mu}_f = 0.5$ (dotted line),  $\tilde{\mu}_f = 0.6$ (dashed line), $\tilde{\mu}_f = 0.8$ (solid line), and $\tilde{\mu}_f = 0.95$ (dash-dotted line). In both panels we have considered $a = 100$~km.  \label{fig:capa}}
\end{figure*}

\begin{figure*}[!p]
\centering
\epsscale{0.99}
\plotone{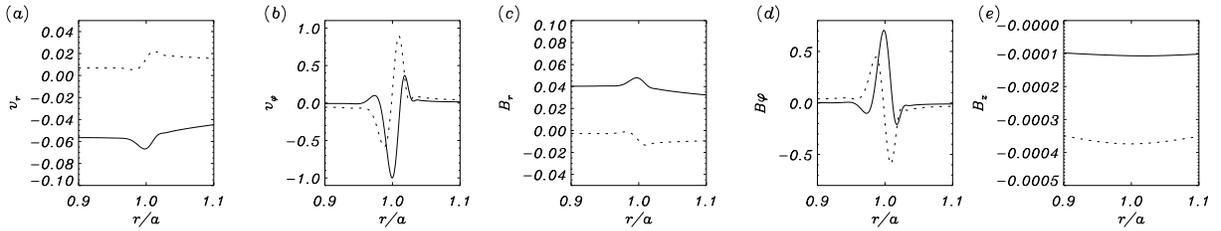}
\caption{Eigenfunctions of perturbations ($a$) $v_r$, ($b$) $v_\varphi$, ($c$) $B_r$,  ($d$) $B_\varphi$, and ($e$) $B_z$ of the kink mode around the resonant point ($r/a \approx 1$) for $l/a = 0.2$, $k_z a = 10^{-2}$, $a = 100$~km, and $\mut_f = 0.8$. The solid line denotes the real part and the dotted line is the imaginary part of the corresponding eigenfunction. Arbitrary units have been used. \label{fig:eigen02}}
\end{figure*}

\begin{figure*}[!p]
\centering
\epsscale{0.99}
\plotone{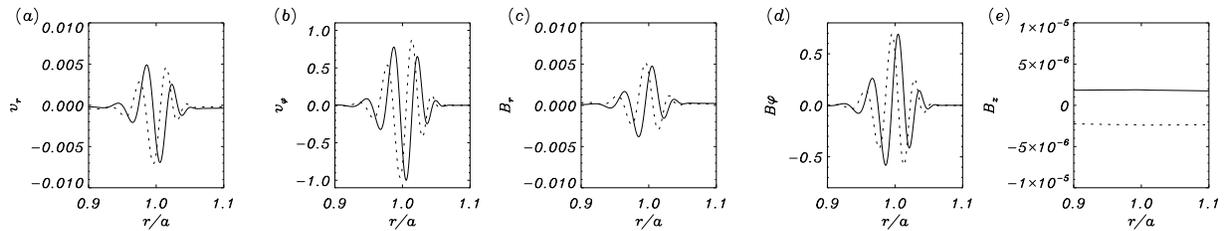}
\caption{Same as Figure~\ref{fig:eigen02} but for $l/a = 0.5$. \label{fig:eigen05}}
\end{figure*}

Hereafter, we include an inhomogeneous transitional layer in the model. This causes the Alfv\'en continuum to be present and so the kink mode is now resonantly coupled to continuum modes. In Figure~\ref{fig:capa}($a$) we assess the effect of the transitional layer on the ratio $\tdp$. For a fixed $\mut_f$, we compare the results corresponding to several values of the transitional layer width. Some relevant differences appear with respect to the case $l/a = 0$. First, we see that $\tdp$ is dramatically reduced for intermediate values of $k_z a$ including the region of typically observed wavelengths. In this region, the ratio $\tdp$ becomes smaller as $l/a$ is increased, a behavior consistent with damping by resonant absorption. This result is confirmed by means of Figures~\ref{fig:eigen02} and \ref{fig:eigen05}, which display the kink mode eigenfunctions close to the resonance point given by Equation~(\ref{eq:resonantpoint}). We see that, with the exception of $B_z$, which is directly proportional to the total pressure perturbation, the other perturbations show small spatial-scale oscillations of large amplitude around the resonance point, i.e., within the so-called dissipative layer. The azimuthal components of both the velocity and magnetic field perturbations have the largest amplitudes because of the coupling to torsional Alfv\'en continuum modes. The thickness of the dissipative layer, $\delta_{\rm A}$, is given by Equation~(51) of \citet{SGH91},
\begin{equation}
 \delta_{\rm A} = \left[ \left| \frac{\omega}{\Delta} \right| \eta  \right]^{1/3}_{r_{\rm A}}. \label{eq:thickdis1}
\end{equation} 
By considering again that in our model the variations of the local Alfv\'en frequency are only due to the variation of the equilibrium density, and approximating the resonant point by $r_{\rm A} \approx a$, we rewrite Equation~(\ref{eq:thickdis1}) as follows,
\begin{equation}
  \delta_{\rm A} = \left[\frac{l}{\pi} \left( \frac{\rho_f + \rho_c}{\rho_f - \rho_c} \right)  \eta  \right]^{1/3}_{r_{\rm A}}. \label{eq:thickdis2}
\end{equation}
We see that $\delta_{\rm A} \sim l^{1/3}$, so the larger the transitional region width, the larger the thickness of the dissipative layer. This is consistent with Figures~\ref{fig:eigen02} and \ref{fig:eigen05}, in which the small spatial-scale oscillations due to the resonance spread over a wider region for $l/a = 0.5$ (Figure~\ref{fig:eigen05}) than for $l/a = 0.2$ (Figure~\ref{fig:eigen02}).

Turning back again to Figure~\ref{fig:capa}($a$), we see that the ratio $\tdp$ for large $k_z a$ is independent of the transitional layer width and coincides with the solution in the absence of transitional layer. The cause of this behavior is that perturbations are essentially confined within the homogeneous part of the thread for large $k_z a$, and therefore the kink mode is mainly governed by the internal plasma conditions. On the other hand, the solution for small $k_z a$ is completely different when $l/a \neq 0$. We note that the inclusion of the inhomogeneous transitional layer (i.e., for $l/a \neq 0$) removes the smaller critical wavenumber, $k_z^{c-}$, and consequently the kink mode exists for very small values of $k_z a$.

\begin{figure*}[!htp]
\centering
\epsscale{0.49}
\plotone{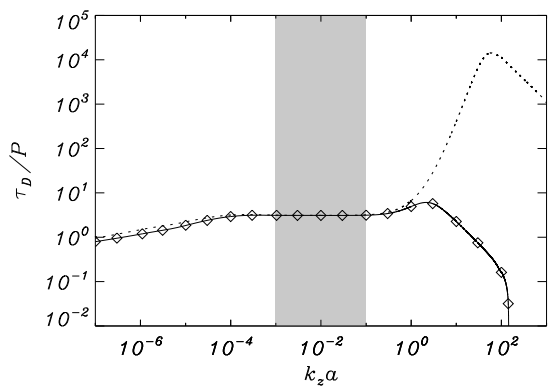}
\caption{Ratio of the damping time to the period of the kink mode as a function of $k_z a$ corresponding to a thread with $a = 100$ ~km and $l/a = 0.2$. The different linestyles represent the results for: partially ionized thread with $\mut_f = 0.8$ considering all the terms in the induction equation (solid line), partially ionized thread with $\mut_f = 0.8$ neglecting Hall's term (symbols), and fully ionized thread (dotted line).  \label{fig:mecha}}
\end{figure*}

We study in Figure~\ref{fig:capa}($b$) the dependence of $\tdp$ on the ionization degree for a fixed $l/a$, and it turns out that the ionization degree is only relevant for large $k_z a$, where the ratio $\tdp$ significantly depends of $\mut_f$. Figure~\ref{fig:mecha} allows us to shed light on this result. There,  we assess the ranges of $k_z a$ where Cowling's and Hall's diffusion dominate. As expected, we find that Hall's diffusion is irrelevant in the whole studied range of $k_z a$, while Cowling's diffusion, caused by the presence of neutrals, dominates the damping for large $k_z a$. In the whole range of relevant wavelengths, resonant absorption is the most efficient damping mechanism, and the damping time is independent of the ionization degree as was analytically predicted in Section~\ref{sec:tt}. On the contrary, we find that ohmic diffusion dominates for very small $k_z a$. It is important noting that although the kink mode is still resonantly coupled to Alfv\'en continuum modes, the damping time related to Ohm's dissipation becomes smaller than that due to resonant absorption, meaning that the kink wave is mainly damped by ohmic diffusion for very small $k_z a$. Since $\eta$ is almost independent on the ionization degree, the ratio $\tdp$ in the region of small $k_z a$ slightly depends on the value of $\mut_f$.

Finally, we compare the full numerical solution with that obtained by solving the analytical dispersion relation in the TB approximation (Equation~(\ref{eq:dispercapa})). For the case $l/a = 0.2$, and $\mut_f = 0.8$, we plot by means of symbols in Figure~\ref{fig:capa}($a$) the result obtained from Equation~(\ref{eq:dispercapa}). We can see a very good agreement between the approximation and the numerical result (solid line) for $k_z a \gtrsim 10^{-4}$, while both solutions do not agree for $k_z a \lesssim 10^{-4}$, which corresponds to the range of $k_z a$ for which ohmic diffusion dominates. Equation~(\ref{eq:dispercapa}) was derived by taking into account the effect of resonant absorption and Cowling's diffusion, but the influence of ohmic diffusion on the damping is not included. For this reason, the approximate solution does not correctly describe the kink mode damping for very small $k_z a$, which corresponds to extremely long and unrealistic wavelengths, while it successfully agrees with the full numerical solution for realistic wavelengths and larger values of $k_z a$.

\section{CONCLUSION}
\label{sec:conclusion}

In this Paper, we have studied the combined effect of resonant absorption and partial ionization on the kink mode damping in a filament thread. Focusing on the results within the observationally relevant range of $k_z a$, we have found that resonant absorption entirely dominates the kink mode damping, with $\tdp < 10$ in agreement with previous investigations \citep[][Paper~I]{arregui}. The obtained damping times are therefore consistent with those reported by the observations. The analytical value of $\tdp$ in the TB and TT limits (Equation~(\ref{eq:tdp})) is a very good approximation to the numerical solution.  None of the other dissipative effects studied here (i.e., ohmic, ambipolar, and Hall's diffusions) is of special relevance in the observed range of wavelengths, and the plasma partial ionization does not affect the mechanism of resonant absorption. The present results reinforce resonant absorption as the best candidate for the damping of filament thread oscillations. 

Some interesting issues not included here might be broached by future investigations. Some of them are, for example, to consider the longitudinal plasma inhomogeneity within the filament thread, to study the spatial damping of propagating waves, and to investigate the excitation and time-dependent evolution of impulsively generated oscillations. 

\acknowledgements{
     RS thanks M. Goossens for reading the manuscript and for giving useful comments. The authors acknowledge the financial support received from the Spanish MICINN, FEDER funds, and the Conselleria d'Economia, Hisenda i Innovaci\'o of the CAIB under Grants No. AYA2006-07637 and PCTIB-2005GC3-03. RS thanks the CAIB for a fellowship.}

\end{document}